\documentclass[a4paper,10pt]{article}
\usepackage{amsmath,amsfonts}
\usepackage{times}
\usepackage[T1]{fontenc}
\usepackage{epsfig}
\usepackage{amssymb}

\begin{document}

\title{\Huge{\bf Compact baby skyrmions}}

\author{C. Adam$^{a)}$, P. Klimas$^{b)}$, J. S\'{a}nchez-Guill\'{e}n$^{a)c)}$, A. Wereszczy\'{n}ski$^{d)}$
       \\
       \\ $^{a)}$Departamento de Fisica de Particulas, Universidad,
       \\ de Santiago and Instituto Galego de Fisica de Altas Enerxias
       \\(IGFAE) E-15782 Santiago de Compostela, Spain
       \\
       \\ $^{b)}$Instituto de Fisica de Sao Carlos,  Universidade de Sao Carlos,
       \\ Caixa Postal 369, CEP 13560-970, Sao Carlos-SP, Brazil
       \\
       \\ $^{c)}$Sabbatical leave at: Departamento de F\'isica Te\'orica, 
       \\ Universidad de Zaragoza, 50009 Zaragoza, Spain
       \\
       \\ $^{d)}$Institute of Physics,  Jagiellonian University,
       \\ Reymonta 4, Krak\'{o}w, Poland}

\maketitle
{\bf keywords:} baby skyrmions, $Q$-balls, compactons, peakons, signum-Gordon model 

\vspace{0.3cm}

\begin{abstract}
For the baby Skyrme model with a specific potential, compacton solutions, i.e.,
configurations with a compact support and parabolic approach to
the vacuum, are derived. Specifically, in the non-topological sector, we find
spinning $Q$-balls and $Q$-shells, as well as peakons. Moreover, we 
obtain compact baby skyrmions with non-trivial topological
charge. All these solutions may form stable
multi-soliton configurations provided they are sufficiently separated.
\end{abstract}
\section{Introduction}
The Skyrme model, i.e., a nonlinear theory of an $SU(2)$ valued matrix
field living in $(3+1)$ Minkowski space-time, plays a prominent
role in high energy physics as a low energy, effective model
of baryonic matter \cite{skyrme}. It provides a non-perturbative
description of nuclei in terms of topological solitons built out
of the primary chiral field. The baryon charge is
identified with the topological index, which also stabilizes
soliton solutions (skyrmions).
\\
The original Skyrme theory can be derived with the help of chiral perturbation
theory, that is, a gradient expansion in terms of the pion field.
In the simplest case it consists of three terms, namely the usual kinetic
term, an expression of fourth order in derivatives known as the Skyrme
term, and (optionally) a non-derivative part called potential. The
Skyrme term is crucial because it allows to circumvent the Derrick
argument for the non-existence of static soliton solutions.
\\
In $(2+1)$ dimensions the Skyrme model possesses its analogue
known as the baby Skyrme model \cite{old}. The main difference
is the field contents. Instead of the chiral field one deals with
a three component unit vector with $S^2$ topology. Then, static
finite energy solutions are maps $ R^2 \cup \{ \infty \} \cong
S^2 \rightarrow S^2$ and, therefore, are characterized by the second
homotopy group of the target space $S^2$, which is nontrivial,
$\pi_2(S^2) = Z$. Similarly to the original theory the baby Skyrme
model contains the kinetic term, a $(2+1)$ dimensional version of
the Skyrme term and a potential. However, now the inclusion of the
potential term is obligatory, as well, in order to circumvent the
Derrick theorem.
\\
Of course, due to the lesser number of dimensions the baby Skyrme
model can serve as a laboratory for the original theory, where
some ideas and methods can be tested. Independently, the baby
Skyrme model found its own applications in condensed matter
physics in the description of the quantum Hall effect \cite{qhe}.
\\
One of the issues widely discussed in the literature is the role of the potential term and its influence for Skyrme solitons (see, e.g. \cite{SkyrmePot}). Since this part of the model does not follow from any rigorous expansion,
its particular form is arbitrary and should be adjusted to a
concrete physical situation. 
For the full 3+1 dimensional Skyrme model, some recent studies of different potentials can be found, e.g., in \cite{PieZa1}. For the baby Skyrme model,
the simplest, one-vacuum cases are
realized by the holomorphic \cite{holom} and the old potential
\cite{old} (potentials with vacuum degeneracy have been studied, e.g., in \cite{other-pot}).
Recently, these potentials have been 
generalized in a natural way by Karliner and Hen to a one-parameter family of
one-vacuum potentials $V=(1-n^3)^s$, with parameter $s \in
[1/2,4]$ \cite{karliner1}, \cite{karliner2}. It has been observed
that the value of the parameter $s$ influences, not only
quantitatively, but also qualitatively properties of topologically
nontrivial solutions, leading to the repulsive or attractive character
of the interaction between baby skyrmions. Moreover, also the
rotational symmetry breaking seems to be governed by the parameter
$s$.\footnote{Double-vacuum potentials have been also considered.
The best-known example is the new Skyrme model which gives
ring-line multi-soliton solutions \cite{new}. For other potentials
of this type see \cite{other}.}
\\
In the present paper we would like to further analyze the case
with $s=1/2$. In fact, this potential has a very peculiar
form with a so-called $V$-shaped singularity at the minimum. It
leads to a completely new qualitatively feature. Namely, the model
possesses compacton solutions.
\\
As is suggested by its name, a compacton is a soliton solution
with compact support. In contrast to standard solitons, compactons
reach the vacuum at a finite distance. Therefore, they do not
possess exponential tails but approach the vacuum in a parabolic, or
generally, power-like manner. In our case, for $s=1/2$, the approach is 
quadratic, i.e., parabolic. We remark that, in principle, there exist compactons
(i.e., a power-like approach to the vacuum) for all $s\in [1/2,1)$, but we
only consider the simplest case of a quadratic approach (i.e., $s=1/2$) in
this paper.  
\\
Originally, compactons have been derived as a class of
one-dimensional solitary waves in generalized versions of the KdV
equation \cite{rosenau1}. Recently, the notion of compact domain
walls (kink) has been extended to topological solitons in Lorentz
invariant field theories. The key ingredient is the inclusion of
so-called $V$-shaped potentials (i.e., potentials which are not
smooth at their minimum \cite{arodz1}). In particular, the left
and right derivatives at the minimum do not vanish and the second
derivative does not exist. The solutions of these theories are therefore typically weak solutions, which are the appropriate solutions for a variational problem, in any case. In practice, this means that there is
no mass scale in the system. An alternative possibility is to
consider K-fields, i.e., fields with a non-standard kinetic term
\cite{comp}, \cite{bazeia1}, \cite{comp br} (outside the context of compactons, the role of K-fields in the formation of topological defects has been studied, e.g., in \cite{Babichev1}). In general, the
appearance of compactons is a result of the mutual relation
between the spatial gradient term and the potential part of the
action \cite{dusuel1}.
\\
As always in non-linear field theory, the extension of one-dimensional
objects to higher dimension is a rather non-trivial issue.
However, higher dimensional compactons have been reported
\cite{rosenau2}, \cite{higher kdv}. Moreover, both $V$-shaped
potentials as well as $K$-fields provide a mechanism for the generation
of compactons also in higher dimensions. For example, one can
mention compact $Q$-balls in the complex signum-Gordon model
\cite{arodz2}, \cite{arodz3} or compact vertices and compact
suspended Hopf shells in some $K$-field models \cite{comp high}.
\\
Our paper is organized as follows. In Section 2 we introduce the version of the baby Skyrme model we study in the rest of this paper. In Section 3 we derive non-topological compact $Q$-balls and $Q$-shells. We also find a special type of solutions called peakons. In Section 4 we derive topological compact baby skyrmions and compare our solutions to the recent results of Karliner and Hen, Refs. \cite{karliner1}, \cite{karliner2}. In Section 5 we discuss a restriction of the baby Skyrme model without the quadratic kinetic term. We re-derive analytic compact baby skyrmions already found in \cite{GP1} and generalize these results. Section 6 contains our conclusions. 
\section{A baby Skyrme model}
In the present paper we focus on a version of the baby Skyrme
model given by the following Lagrangian density
\begin{equation}
L=(\partial_{\mu} \vec{n})^2 -\beta [\partial_{\mu} \vec{n} \times
\partial_{\nu} \vec{n} ]^2- V(\vec{n}), \label{b skyrme}
\end{equation}
where $\vec{n}=(n^1,n^2,n^3)$ is a unit iso-vector living in
$(2+1)$ dimensional Minkowski space-time and $\beta$ is a positive
coupling constant. For our purposes we specify the obligatory
potential term as
\begin{equation}
V=\frac{\lambda}{\sqrt{2}} (1-n^3)^{1/2}, \label{b potential}
\end{equation}
where $\lambda$ is a positive constant. Further, for convenience
reasons, we express the model in terms of a complex scalar field
via the standard stereographic projection
\begin{equation}
\vec{n}=\frac{1}{1+|u|^2} \left( u+\bar{u}, -i ( u-\bar{u}),
1- |u|^2 \right).
\end{equation}
Then,
\begin{equation}
L=4 \frac{u_{\mu}\bar{u}^{\mu}}{(1+|u|^2)^2} - 8 \beta
\frac{(u_{\mu}\bar{u}^{\mu})^2-u_{\mu}^2\bar{u}_{\nu}^2}{(1+|u|^2)^4}-
\lambda \frac{|u|}{\sqrt{1+|u|^2}}.
\end{equation}
The corresponding field equations read
\begin{equation}
 \partial_{\mu} \left( \frac{\mathcal{K^{\mu}}}{(1+|u|^2) ^2} \right)+  \frac{2\bar{u}}{(1+|u|^2)^3} \mathcal{K}_{\mu}
\partial^{\mu} u+
\frac{\lambda}{8} \frac{\bar{u}}{|u|(1+|u|^2)^{3/2}}=0
\end{equation}
and its complex conjugate. Here
\begin{equation}
\mathcal{K}^{\mu}= \bar{u}^{\mu}-4\beta
\frac{(u_{\nu}\bar{u}^{\nu})\bar{u}^{\mu}-\bar{u}_{\nu}^2u^{\mu}}{(1+|u|^2)^2}.
\end{equation}
In the subsequent investigation we assume the following rotationally symmetric Ansatz
\begin{equation}
u=e^{i(\omega t  + n\phi)} f(r), \label{ansatz}
\end{equation}
where $\omega$ is a real parameter and $n \in Z$. Further, $f$ is a non-negative function on the whole interval $r\in [0,\infty )$. Such an Ansatz is
in agreement with the observation that multisoliton solutions of this
baby Skyrme model reveal rotational symmetry, at least for small
values of the model parameters \cite{karliner1}, \cite{karliner2}.
\\
The equations of
motion can then be reduced to an ordinary differential equation for the
shape function $f$
\begin{equation}
-\frac{1}{r} \partial_r \left[ rf' \left( 1+ \frac{8\beta f^2
\left(\frac{n^2}{r^2}-\omega^2 \right)}{(1+f^2)^2}\right) \right]+
f\left( 1+ \frac{8\beta f'^2}{(1+f^2)^2}\right)
\left(\frac{n^2}{r^2}-\omega^2 \right) \nonumber
\end{equation}
\begin{equation}
\hspace*{2.0cm} +\frac{2f}{1+f^2} \left[f'^2-f^2\left(\frac{n^2}{r^2}-\omega^2
\right) \right] + \frac{\lambda}{8} \mbox{sign}(f) \sqrt{1+f^2}=0.
\end{equation}
The total energy is
\begin{equation}
E=\int d^2 x \frac{4}{(1+|u|^2)^2} (u_0\bar{u}_0 + \nabla u \nabla
\bar{u}) + \frac{8\beta}{(1+|u|^2)^4} [(\nabla u \nabla
\bar{u})^2-(\nabla u)^2 (\nabla \bar{u})^2] \nonumber
\end{equation}
\begin{equation}
 +
\frac{8\beta}{(1+|u|^2)^4} [ 2u_0\bar{u}_0 (\nabla u \nabla
\bar{u}) - u_0^2 (\nabla \bar{u})^2 -\bar{u}_0^2 (\nabla u)^2] +
\lambda \frac{|u|}{\sqrt{1+|u|^2}},
\end{equation}
or after inserting our Ansatz
\begin{equation}
\hspace*{-4.0cm} E=2\pi \int_0^{\infty} r dr \left( \frac{4}{(1+f^2)^2}
\left[f'^2+f^2\left(\frac{n^2}{r^2}+ \omega^2 \right) \right]\right.
\nonumber
\end{equation}
\begin{equation}
\hspace*{4.0cm} \left.+
\frac{32 \beta f^2f'^2}{(1+f^2)^4} \left(\frac{n^2}{r^2}+\omega^2
\right) + \frac{\lambda f \mbox{sign} (f)}{\sqrt{1+f^2}} \right).
\end{equation}
Here $\mbox{sign}(f)$ is defined as $\mbox{sign}(f)=1$ for $f>0$ and $\mbox{sign}(f)=0$
for $f=0$.

\section{Nontopological compact Q-balls}
Compact nontopological $Q$-balls are solutions in the form of
Ansatz (\ref{ansatz}), where the profile function starts with a vacuum
value (here simply $f=0$) and tends again to the vacuum at a finite distance.
\\
Before we explicitly construct compactons, we would like to give a general argument for the existence of
compact solutions. First, observe that for $u \rightarrow 0$ our
model reduces to the complex signum-Gordon model
\begin{equation}
\lim_{u \rightarrow 0}  L  = L_{0} \equiv4
 u_{\mu}\bar{u}^{\mu} -
\lambda |u|.
\end{equation}
As has been proved by Arodz et al \cite{arodz2}, \cite{arodz3}, such a model
allows for compact (spinning and non-spinning) Q-balls and Q-shells. Due to the fact that the near
vacuum limit $(u \rightarrow 0)$ is relevant for the existence of
compactons (e.g., this limit allows for checking the power-like
approach to the vacuum, which is essential for compactons) we may
expect that objects of this type should be observed also in the
full model. As we will see below, similar arguments hold also for topological
baby skyrmions (such static solutions cannot exist in the signum-Gordon model 
because of the Derrick theorem).
\\
Next, we want to remark that the typical restrictions which usually hold for the angular frequency $\omega$ for $Q$-balls do not apply in our case. Indeed, in a theory with a normal kinetic term for the complex scalar $u$ and a potential $V(|u|)$,  the following restriction for $\omega$ holds for the case $n=0$ (i.e., non-spinning $Q$-balls) \cite{Coleman1},
\begin{equation}
\mbox{min}(2V/|u|^2)\le \omega^2 \le \mu^2
\end{equation}
where $\mu$ is the resulting mass scale $V''(0)$. For a $V$-shaped potential this mass scale does not exist (it is formally infinite) so the upper bound is shifted to infinity. Further, the lower bound is zero for the signum-Gordon model. As a consequence, the signum-Gordon model has $Q$-ball solutions for all values of $\omega$. On the other hand, we shall see that there does exist a lower limit for possible values of $\omega$ in the full baby Skyrme model studied in the present paper. This lower limit is related to the complicated nonlinear structure of the baby Skyrme model, and there does not seem to exist a simple analytic expression in terms of the model parameters for that lower limit.
\subsection{Expansion at the center}
We plug a series expansion in the vicinity of the origin.
Concretely, we assume for $r \rightarrow 0$ that
\begin{equation}
f(r)=r^k(a_0+a_1r+...).
\end{equation}
From finiteness of the energy we have to assume the following boundary
conditions for $n\not= 0$\footnote{For $n=0$ the situation is slightly more involved, therefore we discuss this case in a separate section, see Section 3.5 below.}
\begin{equation}
f(0) =0, \;\;\; f'(0)=c_0, \;\;\; |c_0| < \infty.
\end{equation}
It is equivalent to the fact that $k \geq 1$. Further,
in order to cancel the potential part which starts with a constant
one gets that $k$ must be an integer less than 3. Thus,
\begin{equation}
k=1,2.
\end{equation}
Let us discuss the first case, i.e., $k=1$. The leading order
expression reads
\begin{equation}
-a_0- 8\beta n^2a_0^3+a_0n^2 + 8\beta n^2 a_0^3=0 \;\;\;
\Rightarrow \;\;\; |n|=1.
\end{equation}
That is, the linear approach to the vacuum in the vicinity of the origin $r=0$ is
indeed observed for the $n=\pm 1$ solutions.\\
In the next possible case, $k=2$, we find at the leading order
\begin{equation}
a_0(4-n^2) - \frac{\lambda}{8}=0.
\end{equation}
One can easily notice that for $n=\pm 2$, the equation leads to a
contradiction. Similarly, for $n \geq 3$ we get that the
profile function is a negative function in a sufficiently close
neighborhood of the origin as $a_0 <0$. This is in 
contradiction to our assumption that $f\ge 0$. Therefore, the
only acceptable value for $n$ is $\pm 1$.
\\
In other words, configurations with $n \geq 2$ cannot nontrivially
start at $r=0$ with a bell-like shape. Instead, such
configurations may form shell-like objects which take the vacuum value $f=0$
inside a certain inner radius $R_1 >0$, as we shall see below.
\subsection{Expansion at the boundary}
We assume a similar series expansion for $r\rightarrow R$, where
$R$ is a finite radial point where the profile function can be
smoothly connected to the vacuum
\begin{equation}
f(r)=A_0(R-r)^s+...
\end{equation}
From the required smoothness of the energy one has to impose
\begin{equation}
f(R)=0, \;\; f'(R)=0,  \;\;\; \Rightarrow \;\;\; s > 1.
\end{equation}
Then, at the leading order we find
\begin{equation}
-A_0 s(s-1)(R-r)^{s-2} +\frac{\lambda}{8}=0,
\end{equation}
with the obvious solution $s=2$. This implies that there is a standard
parabolic approach to the vacuum - a typical feature for
compactons.
\\
As has been proved before, spinning compactons with higher values
of $n$ cannot nontrivially begin at the origin. Instead, one
may consider a solution with two parabolic (compacton) ends at
$R_1$ and $R_2$ forming a shell-like structure. That is, we assume
\begin{equation}
f(r)= \left\{
\begin{array}{cc}
0 & 0 \leq r \leq R_1 \\
g(r) & R_1 \leq r \leq R_2 \\
0 & r \geq R_2
\end{array}
\right.
\end{equation}
with the following boundary conditions
\begin{equation}
\lim_{r \rightarrow R_{1,2}} g(r) = 0, \;\;\; \lim_{r \rightarrow
R_{1,2}} \frac{dg(r)}{dr} = 0,
\end{equation}
\subsection{Numerical solutions}
\begin{figure}[h!]
\begin{center}
\includegraphics[angle=0,width=0.6 \textwidth]{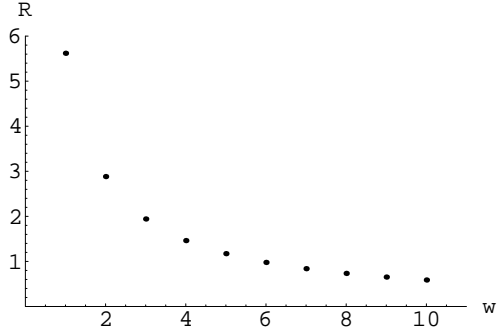}
\caption{Dependence of the size of compact $Q$-balls with $n=1$ on
$\omega$ }\label{n1 R w}
\end{center}
\end{figure}

\begin{figure}[h!]
\begin{center}
\includegraphics[angle=0,width=0.6 \textwidth]{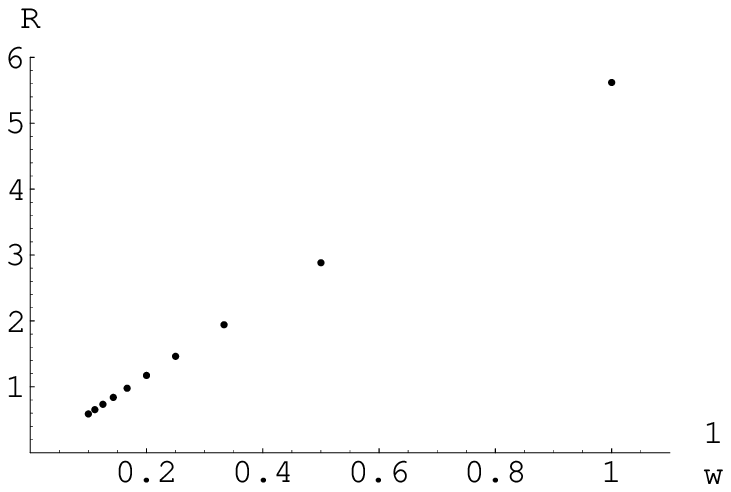}
\caption{Dependence of the size of compact $Q$-balls with $n=1$ on
$1/\omega$ }\label{n1 R w1}
\end{center}
\end{figure}

\begin{figure}[h!]
\begin{center}
\includegraphics[angle=0,width=0.6 \textwidth]{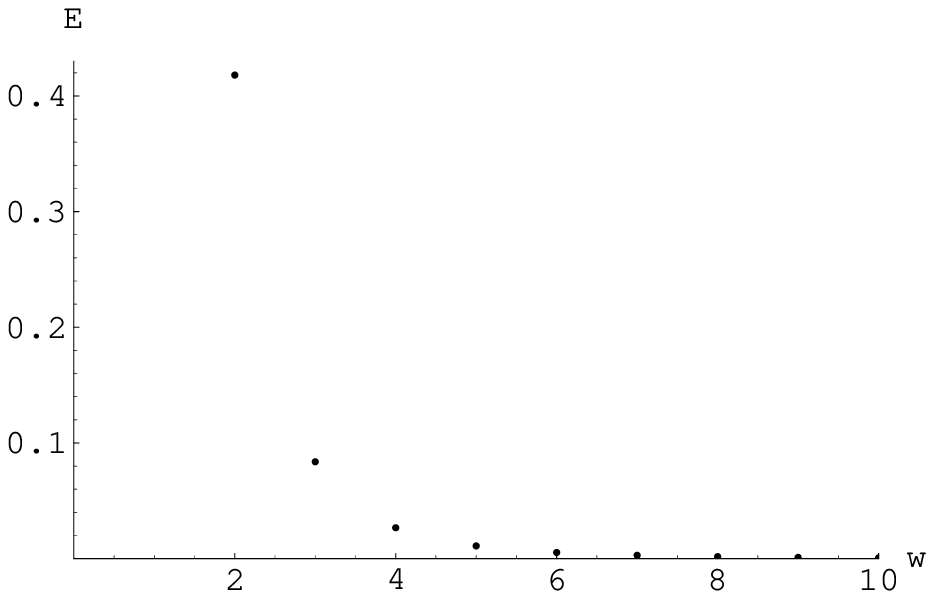}
\caption{Dependence of the energy $E/2\pi$ of compact $Q$-balls with $n=1$
on $\omega$}\label{n1 E w}
\end{center}
\end{figure}

\begin{figure}[h!]
\begin{center}
\includegraphics[angle=0,width=0.6 \textwidth]{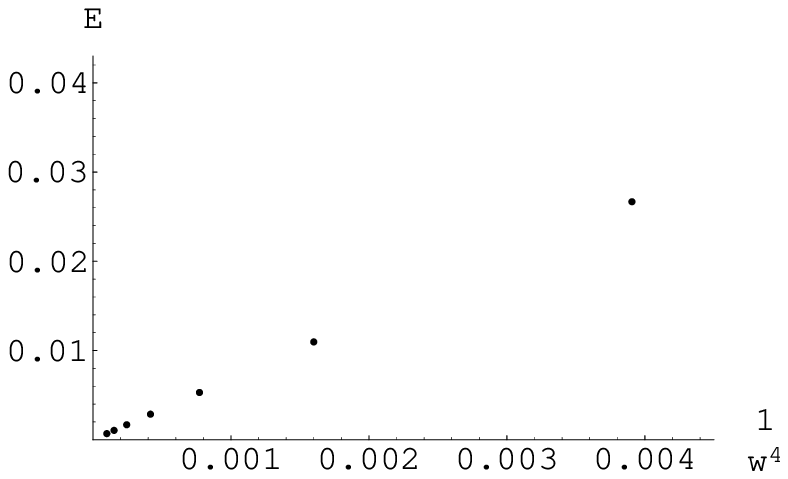}
\caption{Dependence of the energy $E/2\pi$ of compact $Q$-balls with $n=1$
on $1/\omega^4$}\label{n1 E w4}
\end{center}
\end{figure}

Numerical calculations have been performed using the shooting method
starting at a point of the outer boundary of a compact solution and
performing calculation towards $r <R$. Moreover, we assume
$\beta=1$ and $\lambda=1$.
\begin{figure}[h!]
\begin{center}
\includegraphics[angle=0,width=0.45 \textwidth]{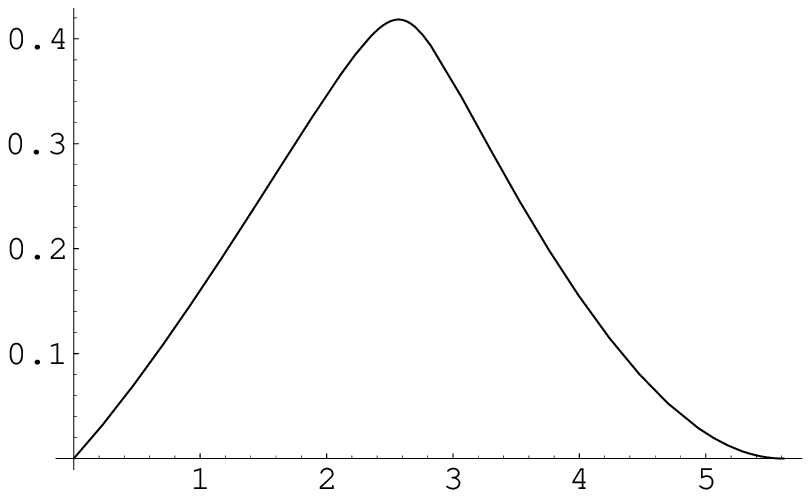}
\includegraphics[angle=0,width=0.45 \textwidth]{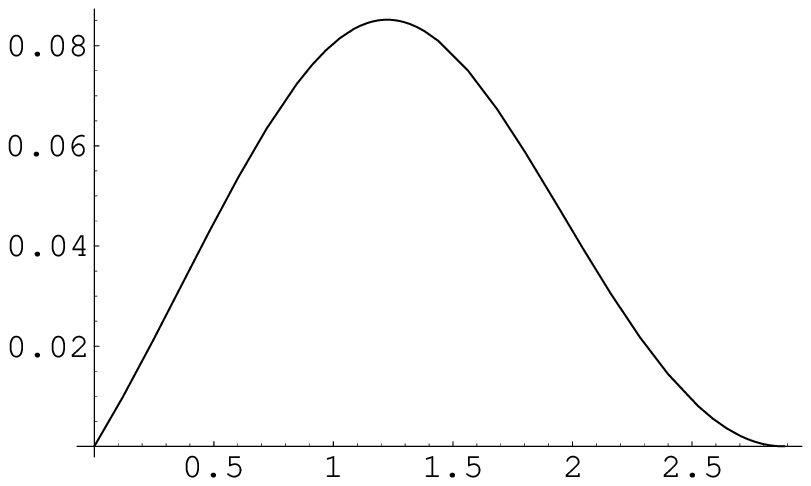}
\includegraphics[angle=0,width=0.45 \textwidth]{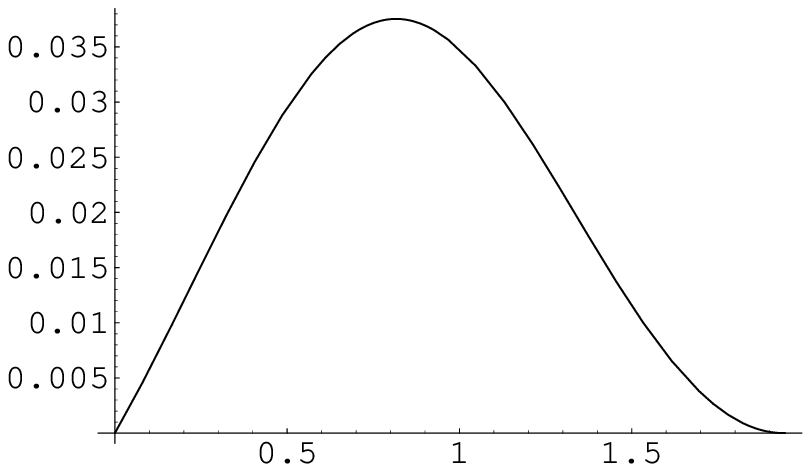}
\includegraphics[angle=0,width=0.45 \textwidth]{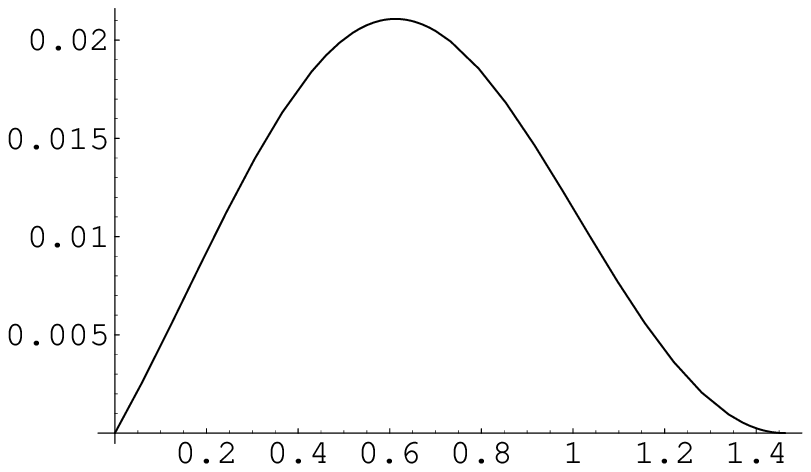}
\includegraphics[angle=0,width=0.45 \textwidth]{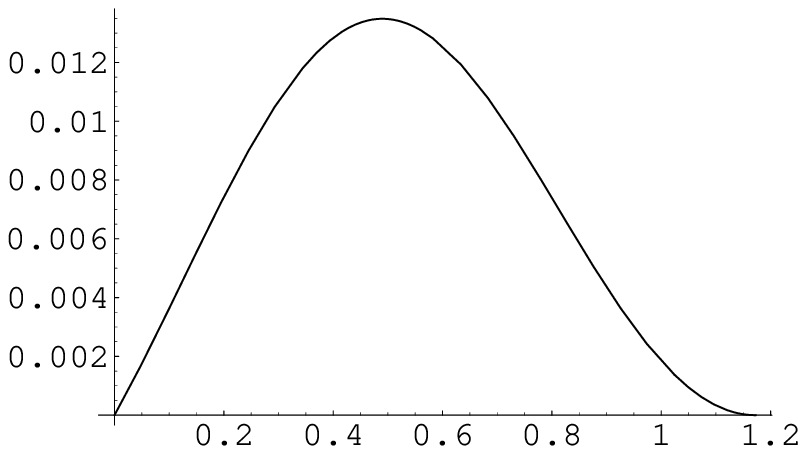}
\includegraphics[angle=0,width=0.45 \textwidth]{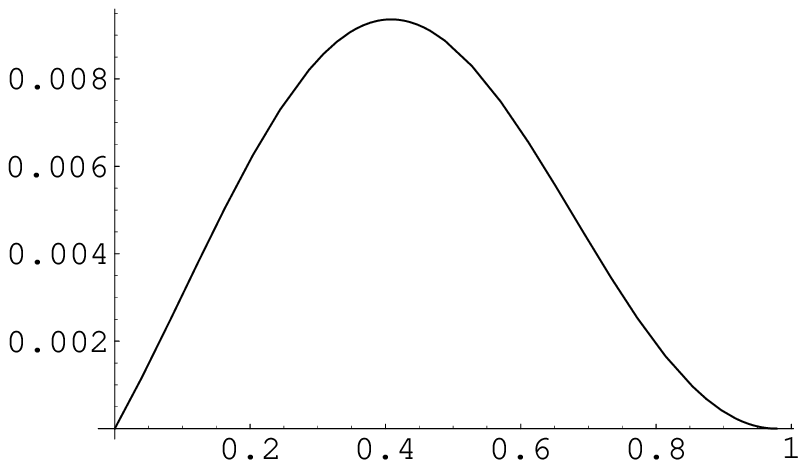}
\includegraphics[angle=0,width=0.45 \textwidth]{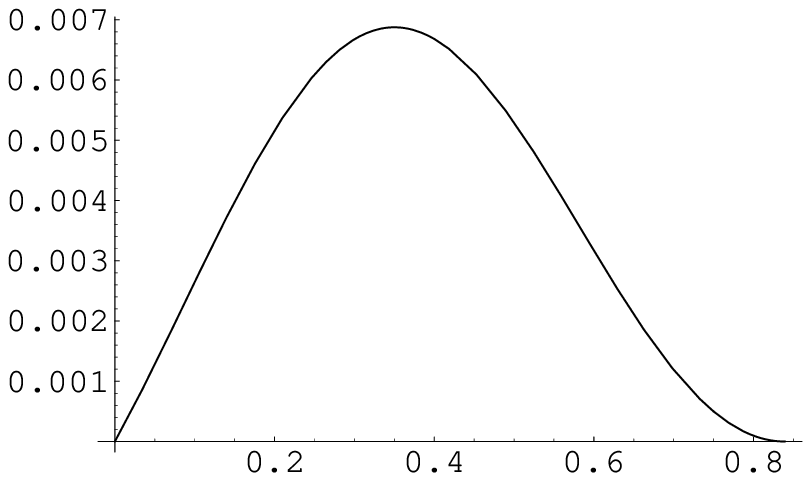}
\includegraphics[angle=0,width=0.45 \textwidth]{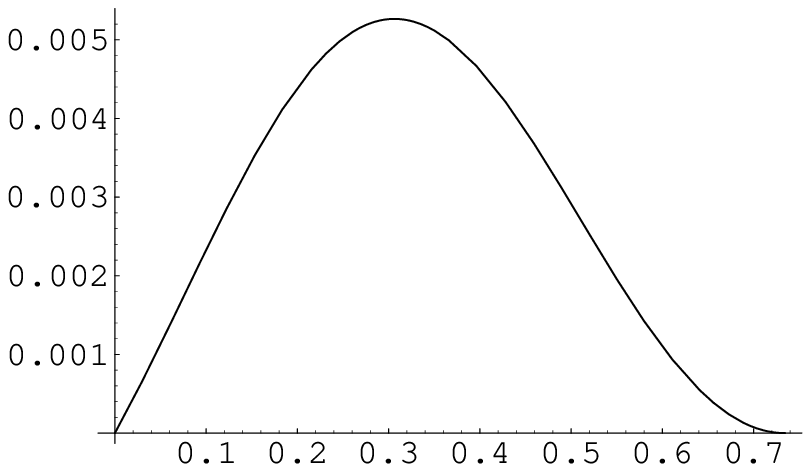}
\includegraphics[angle=0,width=0.45 \textwidth]{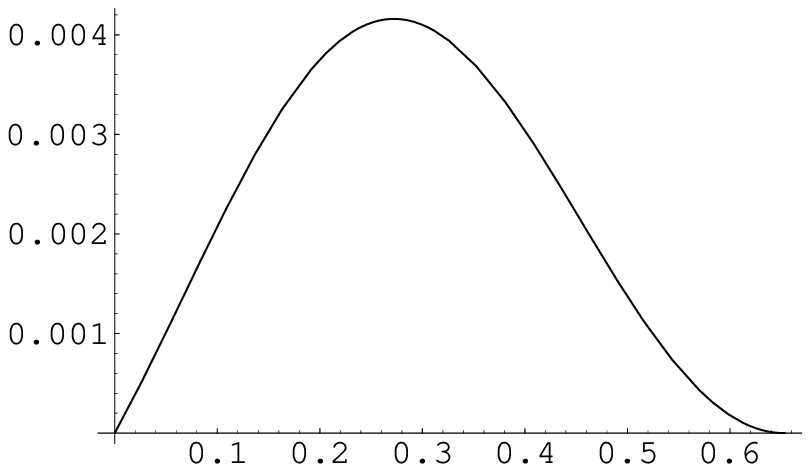}
\includegraphics[angle=0,width=0.45 \textwidth]{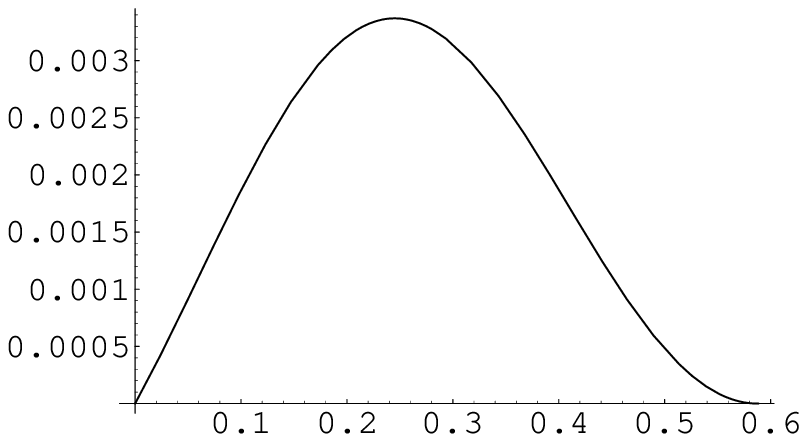}
\caption{Profile function $f$ for non-topological $Q$-balls with
$n=1$ and $\omega=1...10$ }\label{n1}
\end{center}
\end{figure}
\begin{figure}[h!]
\begin{center}
\includegraphics[angle=0,width=0.45 \textwidth]{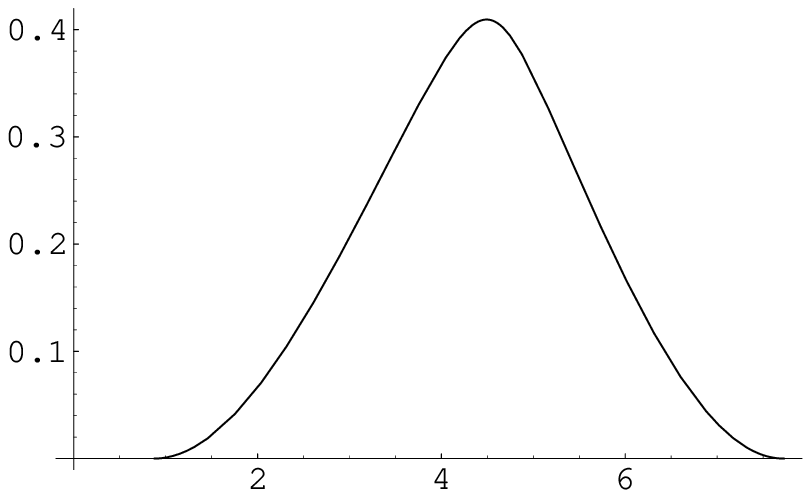}
\includegraphics[angle=0,width=0.45 \textwidth]{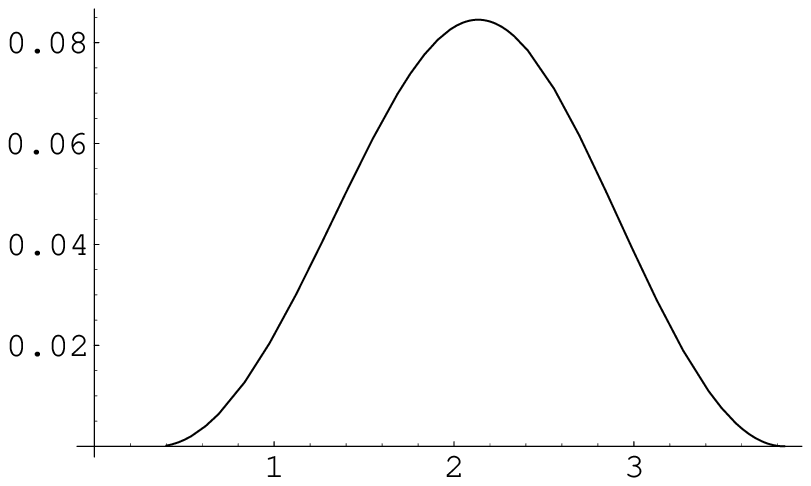}
\includegraphics[angle=0,width=0.45 \textwidth]{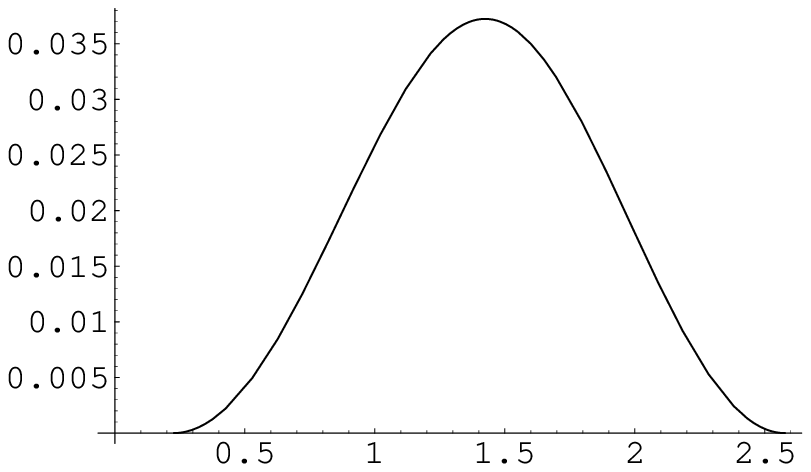}
\includegraphics[angle=0,width=0.45 \textwidth]{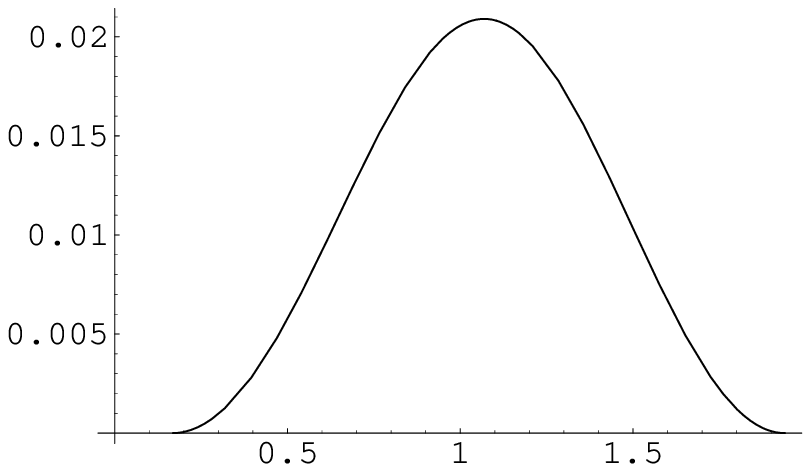}
\includegraphics[angle=0,width=0.45 \textwidth]{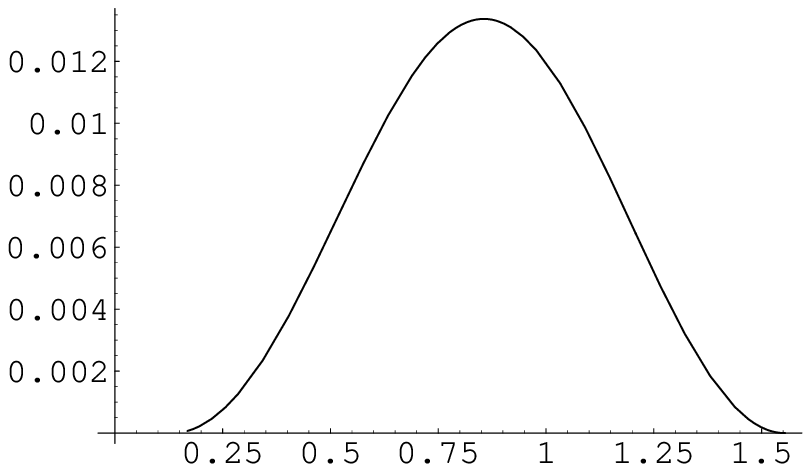}
\includegraphics[angle=0,width=0.45 \textwidth]{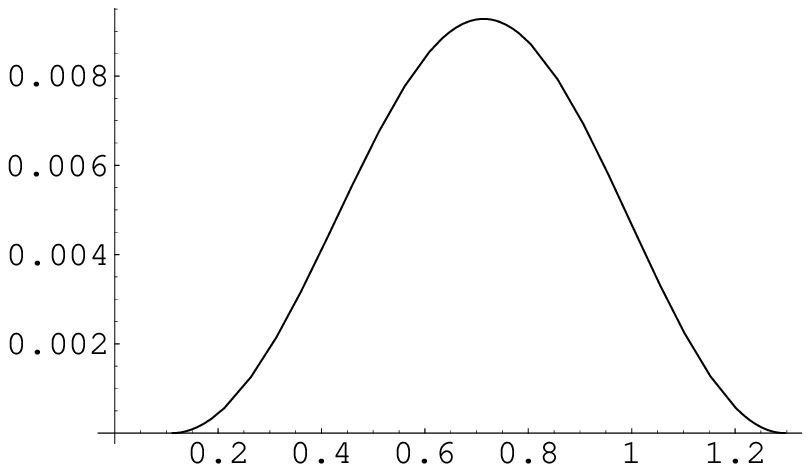}
\includegraphics[angle=0,width=0.45 \textwidth]{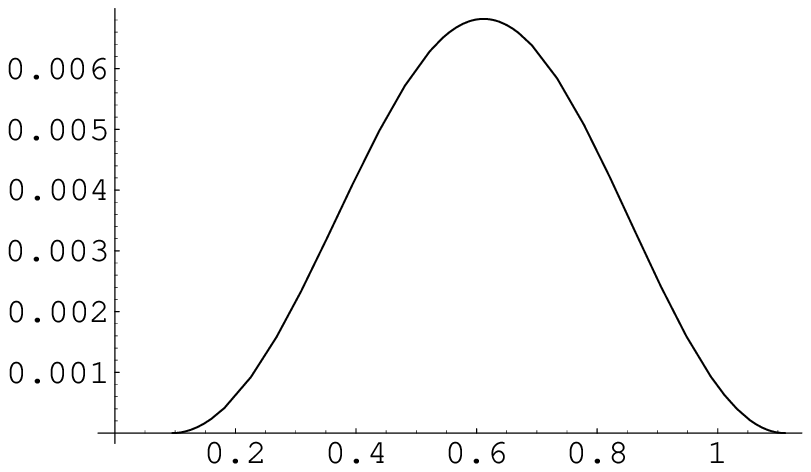}
\includegraphics[angle=0,width=0.45 \textwidth]{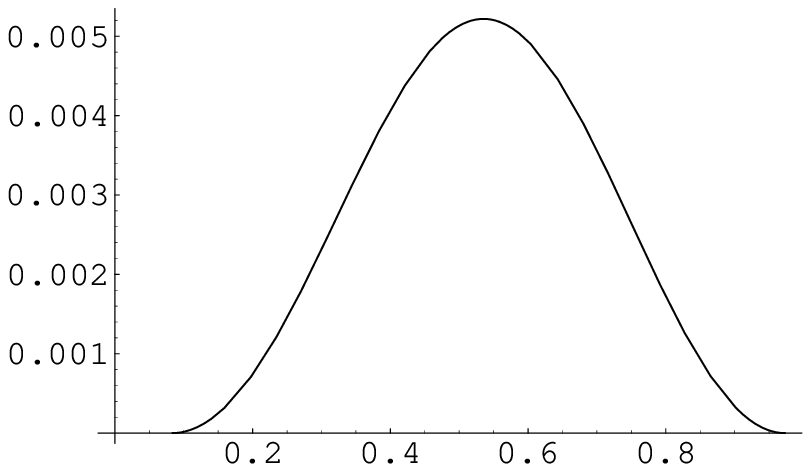}
\includegraphics[angle=0,width=0.45 \textwidth]{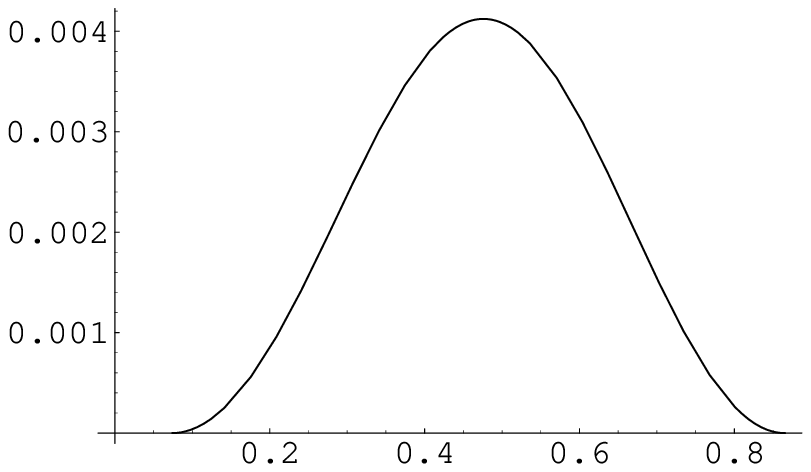}
\includegraphics[angle=0,width=0.45 \textwidth]{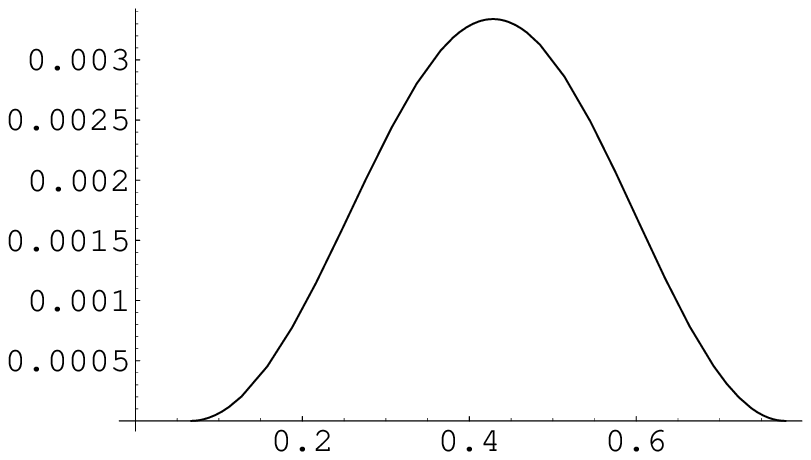}
\caption{Profile function $f$ for non-topological $Q$-shells with
$n=2$ and $\omega=1...10$ }\label{n2}
\end{center}
\end{figure}
Then, for each sufficiently large value of $\omega$ a regular compact configuration
is found for a unique value of $R$. This is completely consistent with a simple count of the
boundary conditions and free parameters. For fixed coupling constants $\lambda$, $\beta$
and $\omega$, in the case of $n=1$ there is one boundary condition, namely $f(0)=0$.
At the same time, there is one free parameter, namely the compacton radius $R$, which has to be fine-tuned accordingly in order to fulfill the boundary condition. For $n\ge 2$, there are two
boundary conditions, namely $f(R_1)=0$ and $f'(R_1)=0$. Further, there are two free parameters, namely the inner and outer compacton radius $R_1$ and $R_2$, which may, again be chosen as to fulfill the boundary conditions.
\\
In Fig. (\ref{n1}) compact $Q$-balls with $n=1$ are presented. One can
recognize a linear approach to the origin as anticipated before. The approach
to the vacuum value $f=0$ at $r=R$ is parabolic, also in agreement with
previous analytical considerations.
\\
As we see, compactons (their size as well as their maximum) become smaller and smaller with the growth
of the spinning frequency $\omega$. To be more specific, the size $R$ of
the compacton is proportional to the inverse of the frequency (see Fig. (\ref{n1 R w}), (\ref{n1 R w1}))
$$ R \sim \frac{1}{\omega}.$$
Similarly, the total energy of compact $Q$-balls also decrees with
the frequency of rotation Fig. (\ref{n1 E w}). One can find that
Fig. (\ref{n1 E w4})
$$E \sim \frac{1}{\omega^4}.$$
Another interesting feature is the fact that the maximum of the profile
function seems to be of a spike-like nature as $\omega$ approaches a certain critical minimum value $\omega \rightarrow \omega_c \lesssim 1$. For
this value of the frequency the numerics breaks down and we are not
able to find compact $Q$ balls with smaller values of $\omega$. We shall say more about this issue in the next section.
\begin{figure}[h!]
\begin{center}
\includegraphics[angle=0,width=0.6 \textwidth]{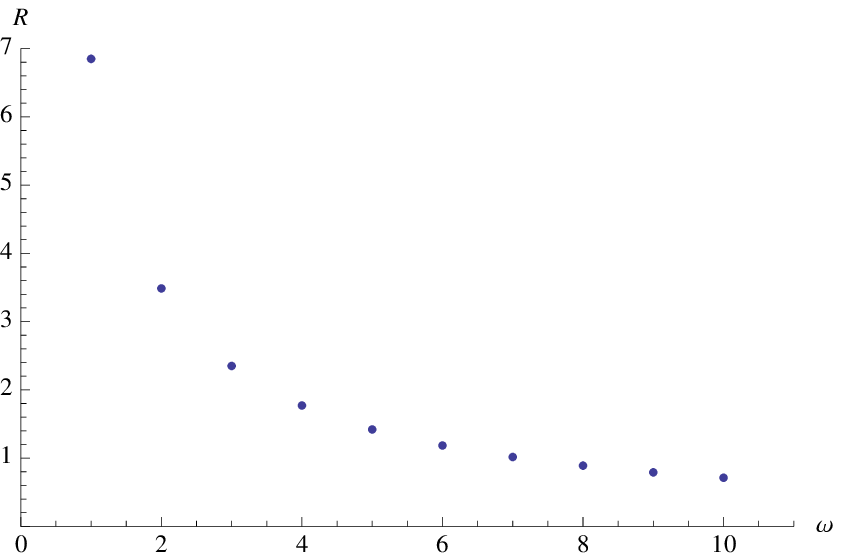}
\caption{Dependence of the size of compact $Q$-balls with $n=2$ on
$\omega$}\label{n2 R w}
\end{center}
\end{figure}
\begin{figure}[h!]
\begin{center}
\includegraphics[angle=0,width=0.4 \textwidth]{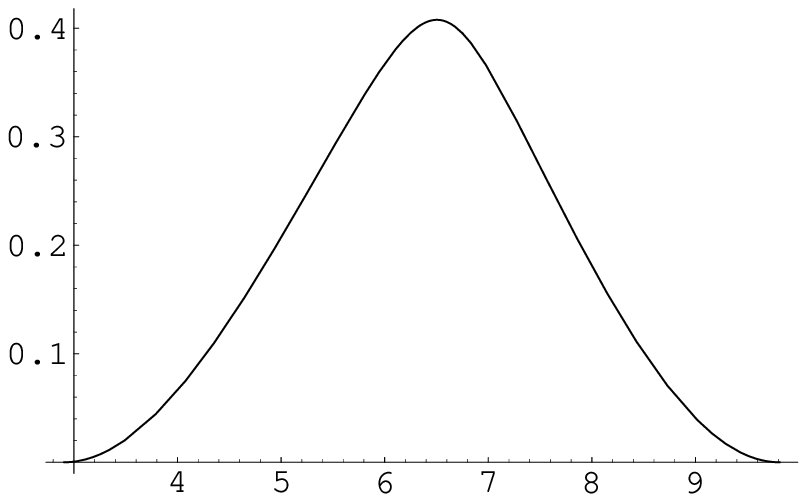}
\includegraphics[angle=0,width=0.4 \textwidth]{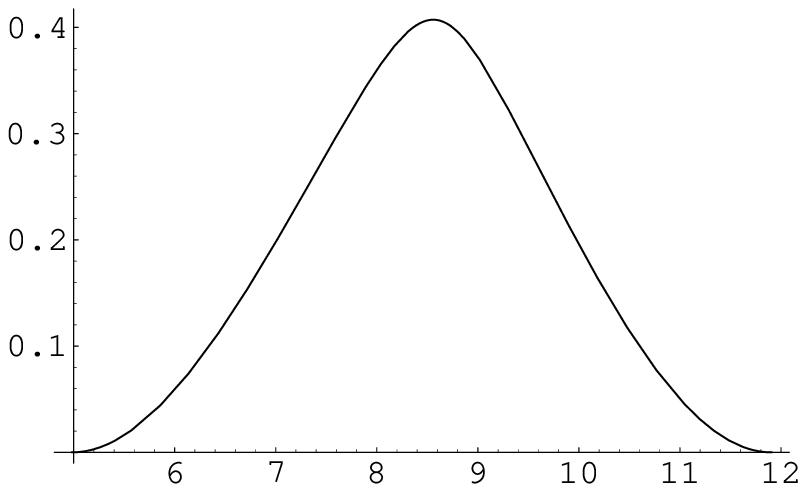}
\includegraphics[angle=0,width=0.4 \textwidth]{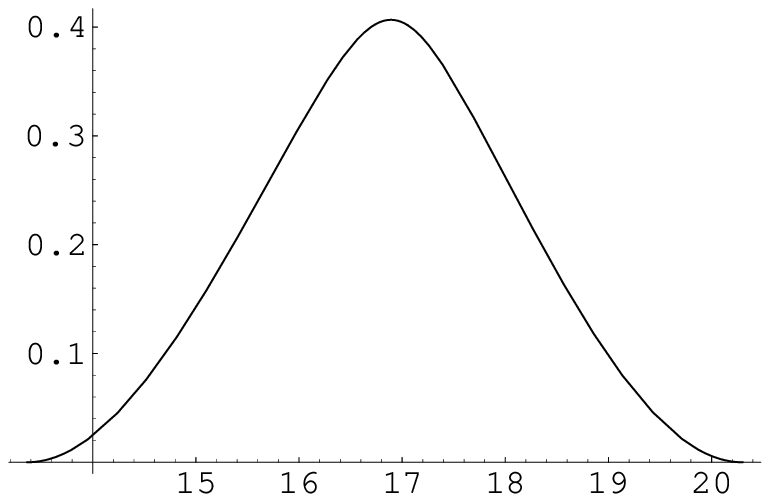}
\includegraphics[angle=0,width=0.4 \textwidth]{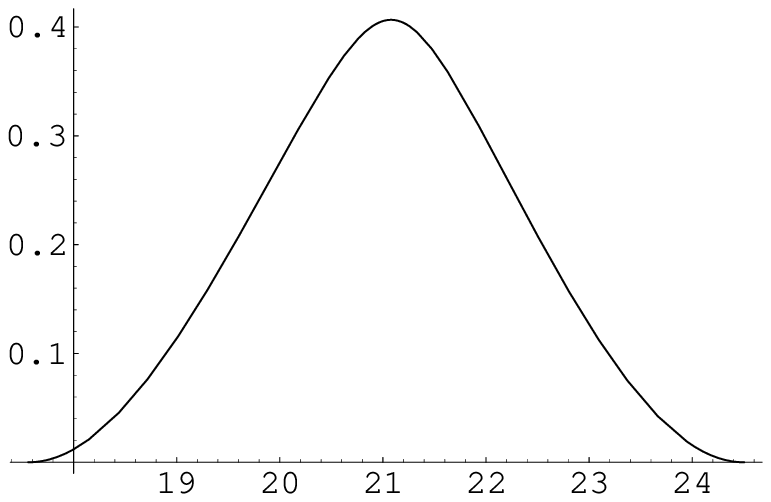}
\caption{Profile function $f$ for non-topological $Q$-balls with
$\omega=1$ and $n=3,4,8,10$ }\label{w1}
\end{center}
\end{figure}
\begin{figure}[h!]
\begin{center}
\includegraphics[angle=0,width=0.6 \textwidth]{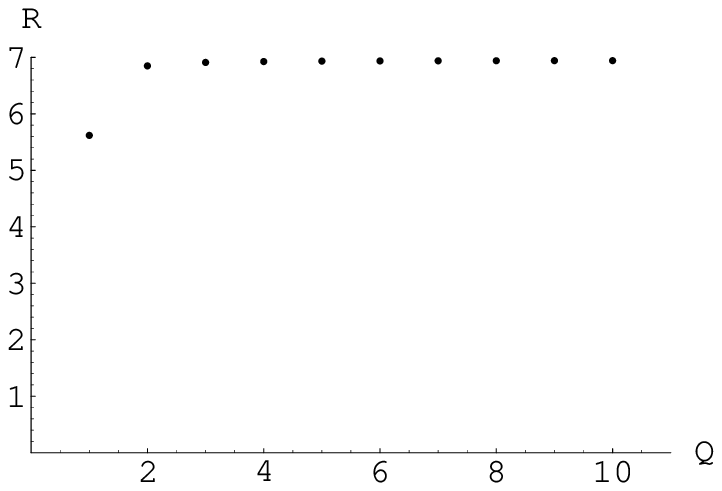}
\caption{Dependence of the size of compact $Q$-balls with
$\omega=1$ on $n$}\label{w1 R n}
\end{center}
\end{figure}
\begin{figure}[h!]
\begin{center}
\includegraphics[angle=0,width=0.6 \textwidth]{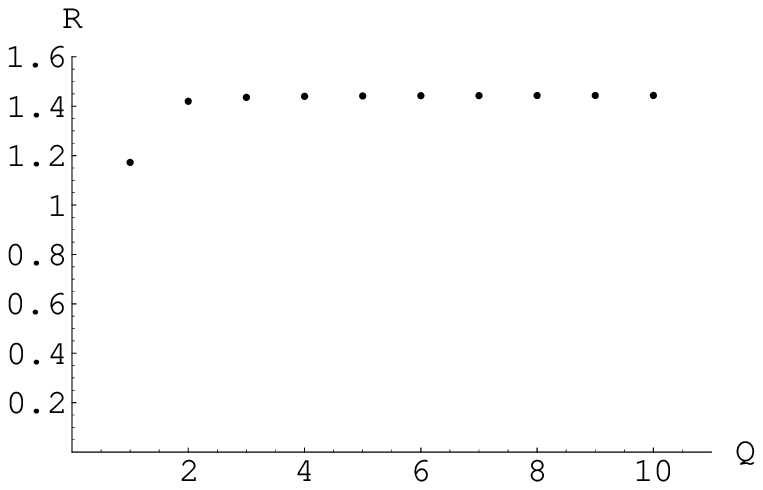}
\caption{Dependence of the size of compact $Q$-balls with
$\omega=5$ on $n$}\label{w5 R n}
\end{center}
\end{figure}
In Fig. (\ref{n2}) we present compact $Q$-shells with $n=2$. In accordance with
the previous analysis compact shells approach the vacuum parabolically at
$R_1$ and $R_2$. As for the $n=1$ solutions, also in this case the size $R=R_2-R_1$ of the compact $Q$-shells is proportional
to $1/\omega$ Fig. (\ref{n2 R w}).
\\
Compact configurations with higher values of $n$ are shown in Fig.
(\ref{w1}) ($\omega =1$). All solutions are of shell shape, where the distance
from the origin grows with $n$. On the other hand, the
size of the non-topological $Q$-compactons quite rapidly tends to a
constant value $R_{\infty}\equiv \lim_{n \to \infty} (R_2 -R_1) =6.94$. 
Of course, the asymptotic size
of nontopological $Q$-compactons scales according to the
previously found law. That is,
$$R_{\infty} \cong \frac{\rho_{\infty}}{\omega}, \quad \rho_{\infty}\cong 6.94 .$$ 
It has been confirmed by numerical checks, see Fig. (\ref{w5 R n}).
\\
One immediately notices that qualitatively the non-topological
$Q$-compactons in the baby Skyrme model behave, to some extent,
similar to the spinning $Q$-balls in the complex signum-Gordon
model. The scaling properties of the energy and size are the same.
Even the asymptotical size of the compactons in the complex
signum-Gordon model takes a similar value and reads $\rho_{\infty
\mbox{sG}}=\frac{4\pi}{\sqrt{3}}\cong 7.25.$
\subsection{Compact peakon}
\begin{figure}[h!]
\begin{center}
\includegraphics[angle=0,width=0.8 \textwidth]{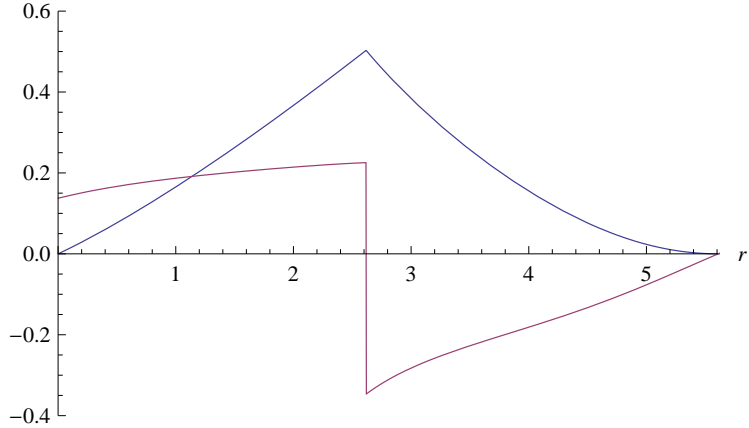}
\caption{$f$ and $f'$ for the compact peakon for $n=1$ and $\omega=0.96$}\label{spike}
\end{center}
\end{figure}

The similarity between nontopological compactons in the baby
Skyrme model and the complex signum-Gordon model originates in the
asymptotic equivalence of these models, i.e., as we noticed it
before, in the fact that for
 $|u| \rightarrow 0$
both Lagrangians takes the same form. As this limit is relevant
for compacton solutions, one can expect that the properties of these
solutions in both theories should be qualitatively similar.
\\
However, due to the strong nonlinearity in the baby Skyrme model,
nontopological compactons do differ from their complex signum-Gordon
counterparts. In fact, spinning compactons in our model exist
only for frequencies which are larger than a certain critical minimum value
$\omega_c$.
\\
A simple explanation of this effect is as follows.
\\
Let us consider the equation of motion at the point $r_m$ where the
profile function takes its maximum $f_m \equiv f(r_m)$. This point must always exist for
a non-topological Q-ball or Q-shell, because the solution must connect to vacua $f=0$. Then,
\begin{equation}
-f''_m \left( 1+ \frac{8\beta f^2_m
\left(\frac{n^2}{r^2_m}-\omega^2
\right)}{(1+f^2_m)^2}\right)+\frac{f_m-f^3_m}{1+f^2_m}
\left(\frac{n^2}{r^2_m}-\omega^2 \right)  +
\frac{\lambda}{8}\sqrt{1+f^2_m}=0.
\end{equation}
where we use that $f'(r=r_m)=0$ and $f''_m \equiv f''(r=r_m)$. It
can be rewritten as
\begin{equation}
\left(\frac{n^2}{r^2_m}-\omega^2 \right)
\left(\frac{f_m -f^3_m}{1+f^2_m} -  f''_m \frac{8\beta
f^2_m}{(1+f^2_m)^2}\right) = f''_m -
\frac{\lambda}{8}\sqrt{1+f^2_m}.
\end{equation}
The right hand side is always negative, as $f''_m < 0$ for a maximum. Analogously,
the second bracket on the l.h.s. is always positive with the additional numerical
information that for $Q$-balls the profile function at its maximum
is less than one, $f_m<1$. Therefore,
\begin{equation}
\omega^2 > \frac{n^2}{r^2_m}
\end{equation}
is a necessary condition for the existence of a solution. Even more can be said if we take into
account that for $\omega = \omega_c$ the  maximum of $f$ at $r=r_m$ is a spike such that
the second derivative $f''_m$ becomes singular. This singularity cannot be cancelled by another term in the e.o.m., therefore it must be multiplied by zero. That is to say, the critical frequency must obey the equation
\begin{equation}
\left( 1+ \frac{8\beta f^2_m
\left(\frac{n^2}{r^2_m}-\omega^2_c
\right)}{(1+f^2_m)^2}\right) =0
\end{equation}
where $r_m$ and $f_m$ are the position of the peak and the value of $f$ at the peak for the critical (peakon) solution. Numerically, this equation holds to a high precision. For $n=1$, e.g., (and for $\beta =1$) we find $\omega_c = 0.9600$, $r_m = 2.618$ and $f_m = 0.5029$ which fulfills the above equation with a precision of about $10^{-4}$.
This ends our proof that the oscillation frequency cannot be
arbitrarily small.
\\
From numerical calculations we found that (for $\lambda =1$, $\beta =1$)
\begin{equation}
\omega_c \approx \left\{
\begin{array}{cl}
0.96 & n=1 \\
0.94 & n=2..10
\end{array}
\right.
\end{equation}
Thus, the minimal frequency is rather $n$-independent. (In
principle, it depends also on the parameter $\beta$ but we do not
discuss this issue in the paper.)
\\
So we found the interesting result that solutions for the minimal frequency are of a
special type, usually referred to as peakons, i.e., solutions with a
jump of the first derivative at the maximum \cite{peakon} - \cite{multipeakon}. Indeed
while approaching the limiting frequency the shape of the
profile function tends to form a spike at the maximum (Fig.
\ref{spike}), and the first derivative is not
continuous, which leads to a singularity in the second
derivative.\footnote{Compact peakons have also been found in the (1+1) dimensional
Hunter-Zheng equation, see \cite{comp peakon}.}
We remark that the peak forms at  the maximum of the function $f$ and, therefore, for values of the field $u$ which are well separated from its vacuum value $u=0$. There is, therefore, no obvious reason which relates the formation of the peakon to the compact nature of the spinning $Q$-ball, and we expect that peakons of an analogous type may form also for spinning $Q$-balls in theories with the standard, exponential approach to the vacuum. This issue certainly  requires some further investigation.

\subsection{Non-spinning compact $Q$-balls}
\begin{figure}[h!]
\begin{center}
\includegraphics[angle=0,width=0.45 \textwidth]{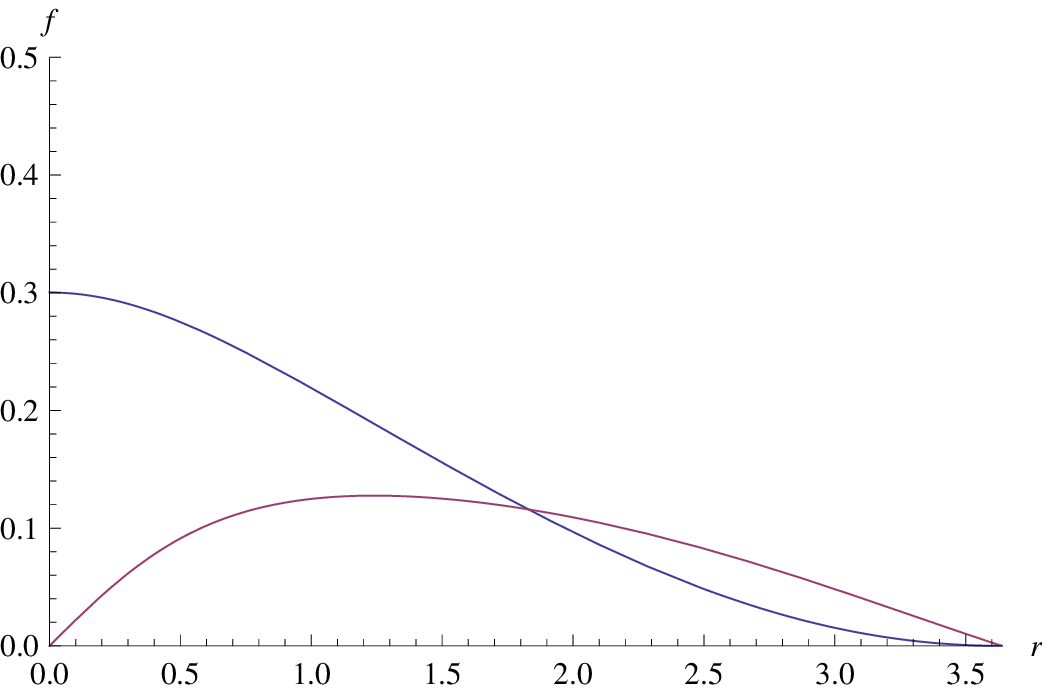}
\includegraphics[angle=0,width=0.45 \textwidth]{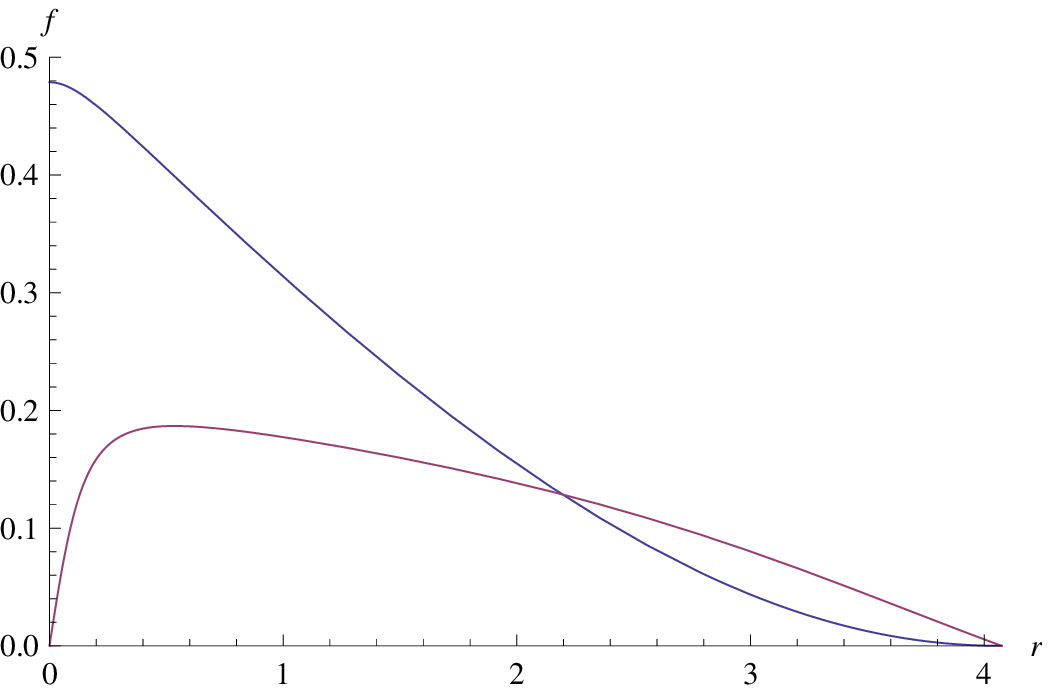}
\caption{$f$ and $-f'$ for the $n=0$ $Q$-balls for the frequencies $\omega =1$ (left) and $\omega = 0.8805$ (right).}\label{fig-Qball-n0}
\end{center}
\end{figure}

We still have to consider the case $n=0$, i.e., the non-spinning, non-topological compact $Q$-ball. A first difference to the case of non-zero $n$ is that the value of the profile function $f(r)$ at the origin $r=0$ is not restricted by either finiteness of the energy or zero topological charge in the case $n=0$. Indeed the topological charge for the spherically symmetric ansatz is 
\begin{equation}
Q=\frac{i}{2\pi} \int d^2 x \frac{\varepsilon^{jk}u_j \bar u_k}{(1+|u|^2)^2} \sim 
n \left[ \frac{1}{1+f^2} \right]_0^\infty
\end{equation}
which is zero for $n\not= 0$ only provided that $f(0)=0$ as well as $f(\infty)=0$, whereas there is no such restriction in the case $n=0$. Inserting the power series ansatz
\begin{equation}
f(r) = a_0 + a_1 r + a_2 r^2 + \ldots
\end{equation}
into the e.o.m. we find that $a_1$ must be zero, whereas $a_2$ is already determined uniquely in terms of the coupling constants and $a_0$ by a linear equation. The expansion at the compacton boundary is identical to the case $n\not= 0$. 
\\
We use again shooting from the compacton boundary for the numerical integration. Here we have one free parameter (the compacton radius $R$) and one boundary condition ($f'(0)=0$), therefore for fixed values of the coupling constants we expect one solution with a fine-tuned value of $R$. Numerically, we find again that solutions only exist for a sufficiently large value of the frequency $\omega$. In addition, the existence of solutions depends also on the values of the additional coupling constants. Specifically we find that for the choices of the coupling constants in the previous sections, namely $\beta =1$ and $\lambda =1$, solutions do not exist for any value of $\omega$. Solutions do exist, for example, for the choice $\beta =1$ and $\lambda = 0.585$, which we assume in the remainder of this section. For these values the critical value of $\omega$ below which solutions do not exist turns out to be
\begin{equation}
\omega_c = 0.8805
\end{equation}
(i.e., a solution exists for $\omega = 0.8805$ but does not exist for $\omega =0.8800$). Further, there does not form any spike at the critical value for $\omega$. It just happens that for $\omega < \omega_c$ the point $f'(0)=0$ cannot be reached by a shooting from the boundary. Instead, $f'$ becomes singular either at $r=0$ or already at some non-zero $r$, depending on the choice of the boundary radius $R$ from which the shooting is done. In Fig. (\ref{fig-Qball-n0}) we show the resulting profile functions $f$ for $\omega =1$ and for the critical value $\omega =0.8805$.   
\\
As is obvious from the results of this and the previous sections, the existence of solutions depends rather crucially not only on the frequency $\omega$, but also on the values of the coupling constants $\lambda$ and $\beta$. Therefore, a scanning of the full parameter space of the theory for solutions might be of some interest and could reveal the existence of different phases in the model. Such a scan is, however, beyond the scope of this article, where we are mainly concerned with demonstrating the existence and the qualitative features of different types of solutions.

\section{Topological compact baby skyrmions}
Firstly, let us remark that the observed similarity of the baby Skyrme model with the complex 
signum-Gordon model of Arodz et al only occurs for
nontopological solutions with a finite maximum value of the
complex field, that is, $|u| \in [0, |u_{\mbox{max}}|]$. 
In the case of configurations with a non-zero value of the
topological charge, $|u|$ may become arbitrarily large, $|u| \in [0,\infty]$, and the geometric
differences of the models start to play an important role. Specifically, there are no static
finite energy solutions in the complex signum-Gordon model, whereas static topological baby skyrmions are known to exist in the model investigated in this paper.
Concretely, topologically nontrivial solutions, i.e., baby skyrmions, may be
found if we specify the following boundary data: at the origin
$f(0)=\infty$ and $f(r=R)=0$, then $u$ (given by the previous Ansatz)
covers the whole $S^2$ and
possesses a nontrivial topology. Notice that as
in the non-topological case, the profile function takes its
vacuum value $f=0$ at a finite
distance.
\subsection{Expansion at the center}
In order to obey the boundary conditions proposed above at the origin
we expand the profile function in the series
\begin{equation}
f(r)=b_0r^{\alpha}+..., \mbox{where} \;\;\; \alpha <0,
\end{equation}
and find at the leading order
\begin{equation}
\frac{8\beta n^2}{b_0^2} \left(
\alpha(\alpha+2)+\alpha^2 \right)r^{-\alpha -4} + (\alpha^2-n^2)
r^{\alpha -2} +\frac{\lambda}{8}r^{\alpha}=0.
\end{equation}
Moreover, the potential part may be also neglected and we get
\begin{equation}
\frac{8\beta n^2}{b_0^2} \left(
\alpha(\alpha+2)+\alpha^2\right)r^{-\alpha -4} + (\alpha^2-n^2)
r^{\alpha -2}=0.
\end{equation}
First we assume that the second term gives the leading
singularity. This happens if $-\alpha-4 > \alpha -2$ i.e., $
\alpha < -1$. Then
\begin{equation}
(\alpha^2-n^2)
r^{\alpha -2}=0.
\end{equation}
with the obvious solution
\begin{equation}
\alpha = - |n|.
\end{equation}
For $\alpha = -1$ both terms are of order $r^{-3}$, but the coefficient of the first term is identically zero and we find
\begin{equation}
 (1-n^2) r^{-3} =0.
\end{equation}
This expression is fulfilled if $n=\pm 1$.
\\
The third possibility occurs if $\alpha > -1$, when the first term is
the leading one. Then,
\begin{equation}
\alpha(\alpha+2)+\alpha^2 =0.
\end{equation}
However, this equation has the only real solutions $\alpha = -1,0$, leading to
a contradiction with our assumption.
\\
To summarize, the leading divergency at the origin reads
\begin{equation}
f(r)=b_0\left( \frac{1}{r} \right)^{|n|}, \;\;\; r \rightarrow 0.
\end{equation}
Let us notice that the expansion at the boundary of compact baby skyrmions
is identical to the non-topological case. Thus,
we may expect the parabolic approach to the vacuum.
Finally, the topological charge $Q$ of these compacton configurations is simply
\begin{equation}
Q=n.
\end{equation} 

\subsection{Numerical solutions}
\subsubsection{Relation to the conventions of Karliner and Hen}
The authors of  \cite{karliner1}, \cite{karliner2} performed a full two-dimensional numerical analysis of topological baby skyrmions for different powers of the potential term. Their main concern was the issue of rotational symmetry breaking, i.e, whether the rotationally symmetric solitons implied by the ansatz of this paper are true minimizers of the energy or just local critical points. They also covered the case of the power $s=1/2$ studied in our paper, although they did not mention the fact that in this case the solitons are, in fact, compactons. As we want to compare their results to ours, in a first step we explain how their and our conventions are related. They use the lagrangian
\begin{equation}
L=\frac{1}{2}(\partial_{\mu} \vec{n})^2 -\frac{\kappa^2}{2} [\partial_{\mu} \vec{n} \times
\partial_{\nu} \vec{n} ]^2- \mu^2 (1-n^3)^{s}
\label{karl-skyrme}
\end{equation}
and for our purpose we shall focus on the case $s=1/2$ in the sequel.
Their normal kinetic term differs from our term by a factor of one-half, therefore even after an identification of all coupling constants our soliton energies will be twice as large as theirs.
Before matching the coupling constants, we want to use the freedom of a scale transformation in order to be able to make simplifying choices for some of them. In fact, the field equations of our model remain unchanged under the following transformation
\begin{equation}
r\to \rho r , \quad \beta \to \frac{\beta}{\rho^2} , \quad \lambda \to \rho^2 \lambda , \quad
\omega \to \rho \omega 
\end{equation}
where $\rho >0$ is a scale factor (further, in the static case $\omega =0$). 
For a general scale factor $\rho$, the relation between the coupling constants of Karliner and Hen and ours  are
\begin{equation}
\mu^2 = \frac{\rho^2 \lambda}{2\sqrt{2}} ,\quad \kappa^2 = \frac{\beta}{\rho^2}.
\end{equation}
Further, Karliner and Hen chose $\mu^2 =0.1$ throughout their numerical analysis, whereas we shall choose $\lambda =1$, which determines the scale factor to $\rho^2 =\sqrt{2}/5$ and, consequently,
\begin{equation}
\beta = \frac{\sqrt{2}}{5} \kappa^2 .
\end{equation}
\subsubsection{Numerical calculations}
\begin{figure}[h!]
\begin{center}
\includegraphics[angle=0,width=0.6 \textwidth]{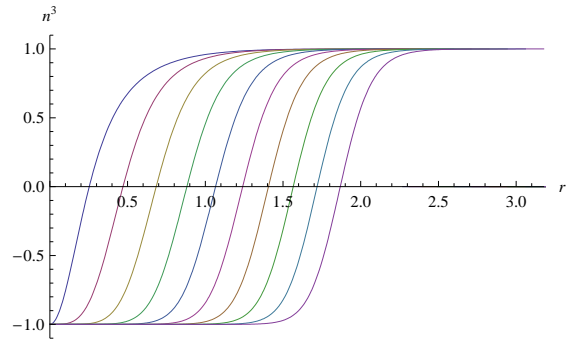}
\caption{$n^3$ component for the compact baby skyrmions with
topological charge $Q=1..10$}\label{top n3}
\end{center}
\end{figure}

\begin{figure}[h!]
\begin{center}
\includegraphics[angle=0,width=0.6 \textwidth]{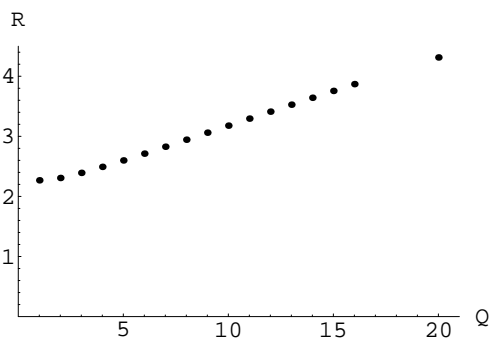}
\caption{Dependence of the size of the compact baby skyrmions on
$Q$}\label{top R}
\end{center}
\end{figure}

Numerical analysis confirms the analytical results presented above. We find
topo\-logical skyr\-mions with a finite radius. The profile
function is singular at the origin, with the expected divergency
$r^{-|n|}$, and parabolically approaches 0 at a certain finite
distance. 
For the parameter choice $\beta = \sqrt{2}/500$ (which corresponds to $\kappa^2 = 0.01$ in the papers of Karliner and Han), we plot 
solutions with topological charge $Q=1..10$ 
 in Fig. (\ref{top n3}) ($n^3$ component of the
iso-vector field) and Fig. (\ref{top E}) (energy density). The
simplest baby skyrmion has a ball shape, with the maximum of the
energy density at the origin. On the other hand, higher compact
skyrmions have a ring-like structure. The location of the maximum
of the energy density grows with the topological charge.
\begin{figure}[h!]
\begin{center}
\includegraphics[angle=0,width=0.6 \textwidth]{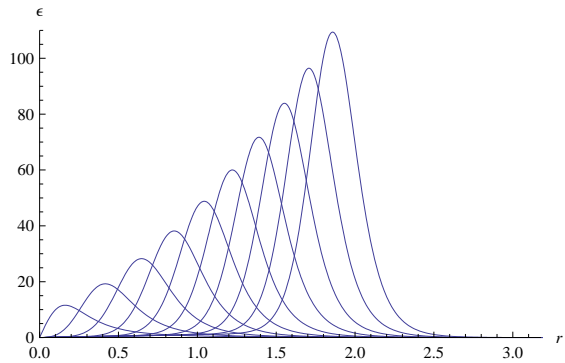}
\caption{Energy density $\epsilon (r)$ (where $E=2\pi \int dr \epsilon (r)$) of the compact baby skyrmions for $\lambda =1$, $\beta =\sqrt{2}/500$, for 
topological charges $Q=1..10$}\label{top E}
\end{center}
\end{figure}
\\
The size of compact baby skyrmions also grows with the topological
charge Fig. (\ref{top R}). In contrast to the non-topological case,
there is no asymptotic (in the sense of the high topological
charge) size of the compactons.
\begin{figure}[h!]
\begin{center}
\includegraphics[angle=0,width=0.6 \textwidth]{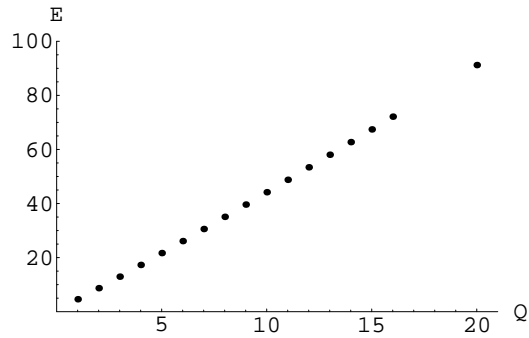}
\caption{Dependence of the energy $E/(2\pi)$ of the compact baby skyrmions on
$Q$}\label{top E}
\end{center}
\end{figure}
\begin{figure}[h!]
\begin{center}
\includegraphics[angle=0,width=0.45 \textwidth]{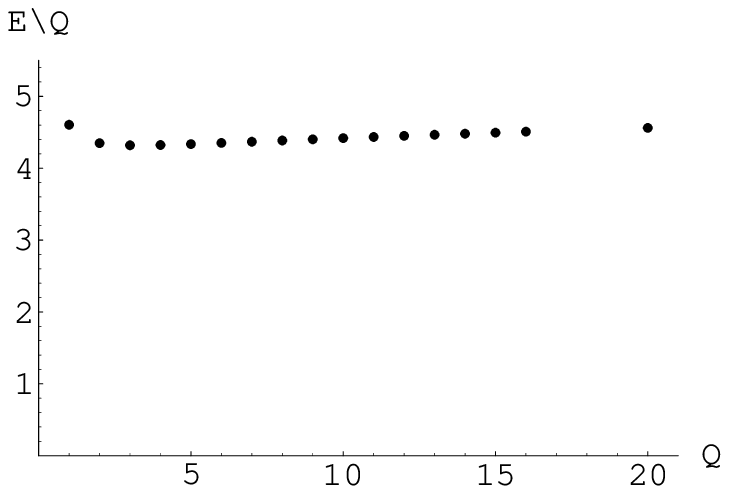}
\includegraphics[angle=0,width=0.45 \textwidth]{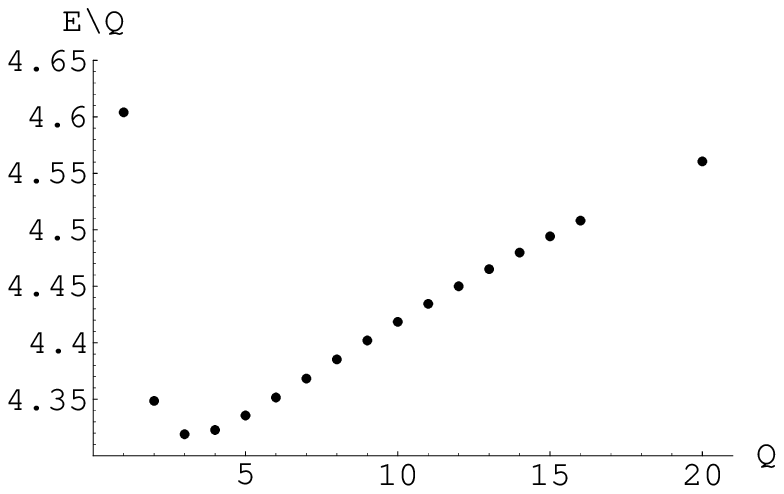}
\caption{$E/(2\pi Q)$ vs. $Q$ of the compact baby
skyrmion}\label{top E/Q}
\end{center}
\end{figure}
\\
Concerning stability, we cannot prove the stability of a topological compacton with our methods, but we can demonstrate its instability. Concretely, a compacton will certainly be unstable if it is heavier than several smaller topological compactons with the same total topological charge. From  this stability criterion we find no sign of instability for the compactons with topological charges from $Q=1$ to $Q=4$. On the other hand, already the $Q=5$ compacton is heavier than one $Q=3$ compacton plus one $Q=2$ compacton, and therefore unstable. All the higher compactons for $Q=6,7,\ldots $ are unstable, as well, see Fig. (\ref{top E/Q}). These findings should be contrasted with the results of Refs. \cite{karliner1} and \cite{karliner2}. There the authors investigated the stability of the spherically symmetric topological baby skyrmions for the same values of the coupling constants for $Q=1,2,3$ and found that all three spherically symmetric solitons are stable. This is fully compatible with our results, and we add the further information that the spherically symmetric baby skyrmions are unstable for $Q=5,6,\ldots$. The stability of the $Q=4$ spherically symmetric baby skyrmion remains an open question. It is not unstable w.r.t. decay into smaller baby skyrmions, but there still could exist a non-symmetric solution with the same topological charge and lesser energy. 
\section{Compactons of Gisiger and Paranjape}
Already in 1996, Gisiger and Paranjape found topological compacton solutions in a version of the baby Skyrme model without the quadratic kinetic term, and with the "old" potential $s=1$,
that is, for the Lagrangian
\begin{equation}
L= -\beta [\partial_{\mu} \vec{n} \times
\partial_{\nu} \vec{n} ]^2- \lambda (1-n^3), \label{GP-skyrme}
\end{equation}
see (\cite{GP1}). Here we want to briefly discuss and generalize their result. For this purpose, we first re-derive their solutions using our field parametrization. For static, spherically symmetric configurations we find for the total energy 
\begin{equation}
E= 2\pi  \int_0^{\infty} r dr \left( \tilde \beta
\frac{ f^2f'^2}{r^2(1+f^2)^4}  + \lambda \frac{ f ^2}{1+f^2} \right)
\end{equation}
where $\tilde \beta \equiv 32 n^2 \beta$. Now, 
we use the freedom to perform a scale transformation $r\to \rho r$ with $\rho^4 = \tilde \beta /\lambda$ to eliminate the coupling constants. We get 
\begin{equation}
E= 2\pi  \sqrt{\tilde \beta \lambda} \int_0^{\infty} r dr \left( 
\frac{ f^2f'^2}{r^2(1+f^2)^4}  +  \frac{ f ^2}{1+f^2} \right) 
\end{equation}
or, after the further transformations 
\begin{equation}
x=\frac{r^2}{2}, \quad 1-g = \frac{1}{1+f^2}
\end{equation}
\begin{equation}
E= 2\pi  \sqrt{\tilde \beta \lambda} \int_0^{\infty} dx (\frac{1}{4}g_x^2 + |g|)
\end{equation}
which is nothing else than the static energy functional of the real signum-Gordon model in 1+1 dimensions. For genuinely 1+1 dimensional compact solitons we would need at least two vacua of the potential, but here we have to take into account the boundary conditions inherited from the two-dimensional problem, namely $f(r=0)=\infty$ and $f(r=R)=0$, $f'(r=R)=0$, which transforms into $g(x=0)=1$ and $g(x=X)=0$, $g_x(x=X)=0$. The corresponding Euler--Lagrange equation is the signum-Gordon equation in one dimension
\begin{equation}
\frac{1}{2}g_{xx} = \mbox{sign} (g)
\end{equation}
and the solution with the right boundary conditions is simply
\begin{equation}
g = (x-1)^2  \quad \mbox{for} \quad x\in [0,1) \, , \quad g=0 \quad \mbox{for} \quad x \ge 1
\end{equation}
with the parabolic approach to the vacuum at $x=X=1$.  This is the static solution of Gisiger and Paranjape. \\
Here, several remarks are appropriate. Firstly, the above result may be easily generalized for
arbitrary potentials $V_s = \lambda (1-n_3)^s$. The energy functional simply changes to
 \begin{equation}
E= 2\pi  \sqrt{\tilde \beta \lambda} \int_0^{\infty} dx (\frac{1}{4}g_x^2 + |g|^s)
\end{equation}
with the Euler--Lagrange equation
\begin{equation}
\frac{1}{2}g_{xx} = s |g|^{s-1}\mbox{sign} (g)
\end{equation}
and the topological compacton solution
\begin{equation}
g = \left( [(2-s)x-1]^2\right)^{\frac{1}{2-s}}  \quad \mbox{for} \quad x\in [0,1) \, , \quad g=0 \quad \mbox{for} \quad x \ge 1 .
\end{equation}
In this case, the approach to the vacuum is still power-like, but no longer quadratic (i.e., parabolic). \\
The second remark concerns the type of compacton we are dealing with in this case (specifically for $s=1$). As the kinetic term is quartic in derivatives, and the potential is standard (quadratic near the minimum), one might believe that this is simply one of the cases of a compacton in a $K$ field theory. This impression is, however, not entirely correct. In a typical $K$ field theory with a quadratic approach to the vacuum, the kinetic term is in fact quartic in the directional derivative perpendicular to the compacton boundary (in the radial direction in our case). That is to say, after a symmetry reduction to an ODE, the energy density near the vacuum typically looks like
$\epsilon \sim f_x^4 + f^2$ where $f$ is a generic field and $x$ the generic independent variable after the symmetry reduction. This behaviour is, however, not possible in the model of Gisiger and Paranjape, due to the anti-symmetry of the quartic Skyrme term. There, the energy density looks more like   $\epsilon \sim f^2f_x^2 + f^2$ near the vacuum, and it may be brought to the signum-Gordon form $\epsilon \sim g_x^2 + |g|$ by the simple map
$g\equiv f^2$ (we remind that even the full model within the spherically symmetric ansatz could be mapped to the signum-Gordon model by a slightly more complicated transformation). These compactons are, therefore, more similar in spirit to the compactons in $V$-shaped potential models than to compactons in $K$ field theories. 
\\
The third remark concerns the possibility to find further solutions in the original model of Gisiger and Paranjape (i.e., for $s=1$) by mapping other ansaetze to versions of the signum-Gordon model. Concretely, we propose the wave-like ansatz
\begin{equation} 
u=f(t,x) e^{i\omega y}
\end{equation}
where $t,x,y$ are the usual rectilinear coordinates in 1+2 dimensional Minkowski space-time. Inserting this ansatz into the Lagrangian of Gisiger and Paranjape we get
\begin{equation}
L=4\omega^2 \frac{f^2}{(1+f^2)^2} (\partial^\mu f)(\partial_\mu f) - \frac{f^2}{1+f^2}
\end{equation}
(here the index $\mu =(0,1)$ refers to 1+1 dimensional Minkowski space) or, after the transformation $1-g = (1+f^2)^{-1}$,
\begin{equation}
L= \omega^2 (\partial^\mu g)(\partial_\mu g) - |g|
\end{equation}
which is again the Lagrangian of the (time-dependent) real signum-Gordon equation in 1+1 dimensions. Now the boundary conditions are like in 1+1 dimensional Minkowski space, so there are no static soliton solutions (because the potential term has only one vacuum). There exist, however, compact time-dependent solutions like, e.g., the oscillon (a bell-shaped profile which oscillates in time), see \cite{arodz4} for details. These oscillons, therefore, provide solutions of the baby Skyrme model of Gisiger and Paranjape.

\section{Conclusions}
It has been the main purpose of the present article to demonstrate the existence of different kinds of compact solutions in the baby Skyrme model with a specific type of potential,
$V=(1-n_3)^{1/2}$. This choice guarantees that the field approaches its vacuum value in a parabolic, i.e., quadratic fashion. For this model we could, indeed, prove the existence of both spinning and non-spinning non-topological compact $Q$-balls and $Q$-shells. As a special case of these, we even found $Q$-balls and $Q$-shells where the profile function develops a peak at its maximum,  so-called peakons. Further, we showed the existence of topological compact baby skyrmions and commented on the relation of our findings in this case with the recent results of Refs. \cite{karliner1}, \cite{karliner2}. Finally, in Section 5 we investigated a slightly different model, namely the baby Skyrme model without the quadratic kinetic term.
In this model, analytical topological compactons have already been found in Ref. \cite{GP1}, 
and we demonstrated that this model may, in fact, be mapped exactly to the signum-Gordon model, which is well-known to support compact solutions.    
\\
A first possible generalization consists in allowing for slightly more general potentials, concretely for the one-parameter family $V_s =(1-n_3)^s$. For $1/2\le s<1$, compact solutions should still exist, although the approach to the vacuum will no longer be quadratic for $s\not= 1/2$. Instead, the field will approach its vacuum value in a power-like fashion with a power different from two. Another type of configurations which we did not investigate in this paper, but which should be quite easy to study with our methods, are topological $Q$-balls, i.e., fields with a non-zero topological charge and with a time dependent phase $\exp (i\omega t)$. Such configurations have been studied in baby Skyrme models, e.g., in \cite{old}, and in the full Skyrme theory, e.g., in \cite{BKS1}.
\\
Next, let us emphasize that due to the finite size of
compacton solutions, the interaction between compactons is of a finite
range type. Two compact baby skyrmions, or non-topological
solutions, may interact only if they are sufficiently close to
each other. Therefore, one may easily construct a multisoliton
static configuration by putting an arbitrary number of compact baby
skyrmions of arbitrary type (i.e., with arbitrary topological
charge) provided that the centrers of the compactons are sufficiently
separated. 
Another consequence of their strictly finite size is that these compact objects will have a scattering behaviour which is probably quite distinct from other models, so the study of scattering is certainly a worthwile enterprise. Other types of time-dependent configurations may be quite interesting, as well. One question is, for instance, whether stable or semi-stable solutions with a time-dependent energy density, like e.g. breathers, exist. Here
we remark that non-topological, time-dependent solutions of the breather type have already been found in the old baby Skyrme model \cite{piette-z1}, therefore we expect that a compact version of these breathers probably exists also in the model studied in the present paper. 
\\
Given the interesting properties which can be found already in the classical theory, the issue of quantization naturally arises. For nonlinear theories like the baby Skyrme model, usually two methods of quantization are employed. In the first case, one identifies a finite number of relevant low energy degrees of freedom ("collective coordinates") which are then quantized by the methods of quantum mechanics. In this approach, therefore, only a finite number of d.o.f. is quantized, which represents a rather crude approximation in some instances. The other method frequently used consists in quantizing small fluctuations about some relevant classical solutions (like solitons or $Q$-balls), that is to say, using perturbation theory about a nontrivial classical configuration. In the case of compactons for $V$-shaped potentials, this second method faces the problem that the contribution of small fluctuations is not small in the vacuum region, essentially because the term $\mbox{sign}(|\delta u|)$ is not small even for very small but nonzero $\delta u$. The use of perturbative methods is, therefore, problematic in theories which support compactons, and nonperturbative quantization methods or alternative expansions probably have to be used from the outset. We conclude that the quantization of compactons, although certainly rewarding given the nice particle-like properties of these objects, is a difficult problem which, in any case, is beyond the scope of the present paper.   
\\
To conclude, we believe that we have unveiled a rich structure of compact solutions in the baby Skyrme model which we investigated in this paper. These findings, and the findings that can be expected from subsequent studies of this and similar models are certainly interesting both for a deeper fundamental understanding of complicated non-linear field theories and for possible applications to planar problems, given the applications the baby Skyrme model and related models have already found in this context.   

\section*{Acknowledgements}
C.A. and J.S.-G. thank MCyT (Spain) and FEDER
(FPA2005-01963), and support from
 Xunta de Galicia (grant PGIDIT06PXIB296182PR and Conselleria de
Educacion). A.W. acknowledges support from the Ministry of
Science and Higher Education of Poland grant N N202 126735
(2008-2010).

\end{document}